\documentstyle[11pt]{article}

\begin{document}

\hfill{Preprint INR 0903a/95}
\begin{center}

{\Large \bf Bubble chain resummation and universality}

\vskip 1cm

{\large \bf N.V. Krasnikov$^a$} and {\large \bf A.A. Pivovarov$^{a,b}$}

{$^a$\it Institute for Nuclear Research of the Russian Academy of Sciences,

Moscow 117312, Russia}

{$^b$ \it Institut fur Theoretische Teilchenphysik,
Universitat Karlsruhe,
D-76128 Karlsruhe,
Germany}

\end{center}
\centerline{\bf Abstract}
We propose a parameterization for contributions
of an infinite set of
diagrams (bubble chains) into physical observables represented
as integrals of running coupling constant
over the finite region in momentum space.
The perturbation theory part of
contributions is rendered well-defined by
the principal value prescrition for treating the Landau pole
while the rest is connected to the gluon condensate
in case of observables allowing the OPE analysis.
The hypothesis of universality, i.e. the use of our
parameterization for non-OPE cases, is discussed.

\vskip 2cm

1. Introduction.

Perturbation theory (PT) in the running coupling constant of strong
interaction is fully understood and well developed technically being
an excellent tool for exploring hard processes in QCD.
However the numerical value of the effective parameter of the
expansion is quite large
\cite{alpha} that forces one to compute several first terms
of perturbative expansion to get a reasonable precision. Experimental
data are permanently getting better while new
higher order terms of
PT expansions appear rarely because of computational
difficulties. This is especially
discouraging because the standard model fits
existing data
well \cite{efl}
and any sensible deflection will be noticeable
at a high level of accuracy only.

Aiming to precise comparison with experiment one should
keep in mind that
PT is asymptotic and cannot provide unlimited accuracy at fixed
$\alpha_s$.
In practice there is no indication yet on any asymptotic character of
known series for physical observables,
the main limitation being the
technical complexity of getting new terms of an expansion.
Few examples are known with several terms of $\alpha_s$
expansion. For $e^+e^-$ annihilation \cite{kat}
and the $\tau$
semileptonic width \cite{bra} the
second subsequent correction of order $\alpha_s^3$ is
comparable with experimental precision in the
$\overline{\rm MS}$
scheme of subtraction.
Terms of higher orders will hardly be available ever:
perturbation theory seems to be saturated \cite{chet}.

And this is good because the computation of higher order terms
is not worth it -- being
asymptotic the PT is also known to be incomplete.
Theoretically recognized source of
nonperturbative effects is the existence of instantons.
Phenomenologically PT looks unsatisfactory as well
-- resonances can not be detected
at any finite order in the coupling constant.

Thus, stuck to QCD as a fundamental model
it is reasonable to go beyond PT from inside by summing particular
infinite sets of diagrams or specific contributions. For simple cases
this has been done on several occasions. The necessity to sum infinite
series appears, for instance, when an analytic continuation from Euclidean to
Minkowskian region is performed \cite{anal,tau}.
Some attention has been also drawn to fermionic bubbles summation
in attempt to go beyond PT \cite{zak}
(as a recent review, see \cite{bal}).

2. Fermionic bubble chains.

The motivation for choosing this particular set of diagrams is
threefold:

i) in $1/N_f$ QED the corrections to photon propagator form a dominant
set of graphs in $N_f\rightarrow\infty$ limit \cite{qed},

ii) numerically the $\beta_0$ expansion
(a QCD counterpart of $1/N_f$ expansion) works well in
$\overline{\rm MS}$ subtraction scheme for all known cases \cite{be0},

iii) they constitute a well
defined set of graphs that can be summed.

There are different and quite serious objections to above motivation
but the third point is decisive; one wants to find diagrams that can
be summed in a constructive way.

Trying to keep this approach alive one could put forward some other
foundations for fermionic bubbles chains at least as possible
indicator of leading diagrams. Thus, skeleton expansions \cite{skelet}
or
Schwinger-Dyson equations \cite{schdys}
could lead in simple cases to the very same
set of leading graphs in specific approximations. Early efforts to
solve the confinement problem using a modified gluon propagator \cite{arb}
lie in the
same bed.

3. Integral at small momenta.

Whatever the underlying motivation is there are observables that can
be represented on a general ground as an integral of running coupling
including an infrared region.
Examples have been discussed in the literature, we mention three of them
$e^+e^-$ annihilation \cite{zak},
pole mass of heavy quark \cite{big}
and event shape \cite{man,web,akh}.
Being infrared safe these
observables can be formally written
down as integrals over
the running coupling that encounters the Landau pole present in the PT
expression for $\alpha_s(t)$ at small $t$.
Therefore
their expansion in $\alpha_s(Q)$,
where $Q$ is a large scale involved in the
process, generates a factorial growth of PT coefficients.
The emerging series is sign-definite and cannot be
summed with Borel technique.
Thus for a number of observables the representation of the form
\begin{equation}
\int_0^{Q^2}\alpha_s(t) \omega(t) dt
  \label{int}
\end{equation}
can be formally obtained.
The corresponding PT expression reads
$$
\alpha_s(Q^2)\int_0^{Q^2} \omega(t) dt
$$
and the difference between these two
formulae reflects the result of ``bubble
summation improvement''.
The weight function  $\omega(t)$ depends on kinematical invariants
that are of order $Q^2$.
The problem reduces to a proper definition of the
integral (\ref{int})
that
diverges due to the Landau pole if the perturbative running coupling
is used in the infrared region of integration.

Note that high orders of perturbation theory are normally
analyzed
using the Borel transform in the coupling constant. Then nonsummable
pieces reveal themselves as poles in the complex plane
of a Borel parameter. The
growth of coefficients
in fact can be much slower
than factorial -- it is enough for the series
to be sign-definite to
be nonsummable and to produce a singularity.
Such singularities are called renormalons with specification from what
region they come -- IR or UV.
So, factorial divergences are called
renormalons though singularities can be connected with other types of
large $n$ behavior as well.
In practice the simple way to generate coefficients
at any $n$ is to use the running coupling and the presence of Landau
ghost is equivalent to existence of renormalons
as factorial divergences of PT series.
One could say that renormalization group sums logs
while the proper definition of the behavior of
the coupling constant at small momenta sums bubble chains.

Qualitatively such chains are expected to produce
nonperturbative pieces because
$\alpha_s(Q^2)$ and $\alpha_s(t)$ are not connected perturbatively for
$t\rightarrow 0$ and new terms are plausible, {\it i.e.}
a new parameterization, different from purely perturbative,
can enter the game.

Dealing with factorial divergence in quantum field theory
(rewritten as singularities
in the Borel plane) at present state of art means to define how to treat
the Landau ghost. Nevertheless in the literature the interpretation of
the pole in the Borel plane is widely used
\cite{gru}. For instance, the
principal value prescription is very popular \cite{neu}.
In fact, the
principal value prescription in the Borel plane is equivalent to the
principal value prescription in momentum space at least at
the level
of one
loop. Namely,
a Borel represented observable of the form
$$
PV\int_0^\infty {e^{-{t\over \alpha}}dt \over 1-t}
$$
after the change of the variable $t\rightarrow \alpha log(Q^2/t)$
goes to
$$
PV\int_0^{Q^2}\alpha(t)dt=
lim_{\epsilon \rightarrow 0}\left(\int_0^{\Lambda^2-\epsilon}
+\int_{\Lambda^2+\epsilon}^{Q^2}\right){dt\over log({t\over \Lambda^2})}
$$
where we defined $\alpha(t)=\beta_0\alpha_s(t)=1/log(t/\Lambda^2)$.
The last prescription is purely perturbative however. Namely, consider
the integral with $\omega(t)=1$, other functions do not change things
qualitatively. Then
\begin{equation}
PV\int_0^{Q^2}\alpha(t)dt=
\int_{Q^2_0}^{Q^2}\alpha(t)dt=
\alpha(Q^2)\int_{Q^2_0}^{Q^2}\sum_{n=0}^\infty
\left(\alpha(Q^2)log(Q^2/t)\right)^ndt
\label{pvpt}
\end{equation}
and
$$
PV\int_0^{Q^2_0}\alpha(t)dt=0,\quad Q^2_0=\Lambda^2*1.45...
$$
The position of the zero $Q^2_0$
($Q^2_0=\Lambda^2*1.45...$)  of the function
$$
f_\omega(Q^2)=PV\int_0^{Q^2}\alpha(t)\omega(t)dt
$$
depends on the weight function $\omega(t)$
but the zero itself
does exist for any function of the same sign at small $t$.
The series (\ref{pvpt}) is a convergent PT series.
Indeed, inside the integration region we have
$$
|\alpha(Q^2)log(t/Q^2)|<|\alpha(Q^2)log({1.45\Lambda^2\over Q^2})|
=1-{log 1.45\over log {Q^2\over\Lambda^2}}<1.
$$
and the integrand converges homogeneously that allows one to integrate
it getting again the convergent series.
It is clear what happened -- the most interesting region has been
thrown away completely by choosing the PV prescription.

Still this is a definition of the perturbation series.
The coefficients of $\alpha_s(Q)$ are not just numbers but are
functions of $Q^2$ and contain powers of $\Lambda/Q$ as well.
In fact any cut will introduce another scale that renders the
coefficients to become some functions of the ratio of those scales,
in spirit though
the expansion remains purely perturbative. In this sense the
cut with principal value prescription is minimal because it
does not introduce any new scales.
The integral
in eq. (\ref{pvpt}) can be computed explicitly
$$
\int_{Q^2_0}^{Q^2}
\left(log(Q^2/t)\right)^ndt
=Q^2\Gamma(n+1,log(Q^2/Q^2_0))
$$
where $\Gamma(n,z)$ is an incomplete $\Gamma$-function
\begin{equation}
\Gamma(n+1,z)=\int_0^ze^{-t}t^ndt=n!-\int_z^\infty e^{-t}t^ndt
\label{gamma}
\end{equation}
that reduces to elementary functions for our particular case
though.
The last term in eq.~(\ref{gamma}) behaves asymptotically at
large $z$ as $exp(-z)$, or $exp(-1/\alpha(Q^*)$, where
$Q^*=Q/\sqrt{1.45}$. Thus, any coefficient function of the new
perturbative expansion (\ref{pvpt}) contains ``nonperturbative
terms'' (see, also \cite{bound}).

Numerically this recipe is valid meaning that two expressions
$$
PV\int_0^{Q^2}\alpha(t)\omega(t)dt
\quad {\rm and}\quad
\alpha(Q^2)\int_0^{Q^2}\omega(t)dt
$$
do differ.
As we have shown however this difference can be accounted for
perturbatively.
We find
$$
{PV\int_0^{Q^2}\alpha(t)\omega(t)dt\over
\alpha(Q^2)\int_0^{Q^2}\omega(t)dt}=1.39,~1.27,~1.23
$$
for $\omega(t)=1$ and $Q^2/\Lambda^2=100,~500,~1000$.
The convergence of the series looks like (for
$Q^2/\Lambda^2=100$)
$$
1.39=1+0.201+0.075+0.038+0.015+\ldots
$$
or in terms of $\alpha(100)$
$$
1+
0.92\alpha(100)+1.59\alpha(100)^2+3.67\alpha(100)^3+10\alpha(100)^4
+30\alpha(100)^5+\ldots
$$
instead of factorial growth
$$
1+
1!\alpha(100)+2!\alpha(100)^2+3!\alpha(100)^3+4!\alpha(100)^4
+5!\alpha(100)^5+\ldots
$$
The convergence is very slow. To reach a reasonable accuracy
one needs almost as many terms
as one could keep for the initial asymptotic series to get
the best approximation. In this sense it imitates the
asymptotic series very closely.

The convergence can be essentially improved by choosing
some other expansion parameter that reduces to the change
of the scale $\alpha(100)\rightarrow \alpha(100/\xi),~\xi>1$.
The BLM choice \cite{blm}
corresponds to vanishing of the first order
correction;
other optimization criteria lead to their own choice of
the scale.
On the whole, however, all these choices are perturbative in
the sense that they give convergent series in the coupling
normalized at some high scale.
In practice convergence still
can be slow and one is forced to
use the integral formula but in principle this is a
possible minimal way to define the perturbative series.
Thus, the PV prescription
does not create nonperturbative terms though
it allows one
to sum up some PT corrections in a closed form. Do the real
nonperturbative terms exist in the chain?

Look at the running coupling more carefully.
The PV prescription suffers of being nonpositively defined at small
momenta
that can contradict some general properties of quantum
field theory \cite{krasn}. For instance,
in the
case of current correlators the spectral density must be positive.

4. Extrapolation to infrared region.

We give several motivations and models for behavior of the coupling
constant at small momenta keeping positivity \cite{bub}.

First we use the freedom of choosing the renormalization scheme
and coupling definition eventually. Consider the $e^+e^-$
annihilation. Corresponding $D(Q^2)$ function is known up to the
$\alpha_s^3$ order. Redefining $\alpha_s$ in a scheme without higher
order corrections we find
$$
D(Q^2)=1+{\tilde{\alpha}_s(Q^2)\over \pi}.
$$
Evolution of the new charge
$\tilde{\alpha_s}(Q^2)$
is governed by a new $\beta$ function
$\tilde{\beta}(\tilde{\alpha_s})$
that has an IR fixed point and the effective charge is frozen at
small momenta.
The same conclusion has been recently obtained in \cite{pms}
after
analyzing the $e^+e^-$ annihilation cross-section
$R(s)$ within the principle-of-minimal-sensitivity approach
\cite{pms0}.
Even without this reference one can choose a special $\beta$ function
providing a smooth infrared behavior for the coupling.
For instance,
\begin{equation}
\beta(\alpha)=-{\alpha^2\over
  1+\kappa\alpha^2},\quad\kappa>0
  \label{betamod}
\end{equation}
where
$
\beta^{as}(\alpha)=-\alpha^2+\ldots
$
\cite{bub}.
Note, that this extrapolation is based on PT consideration
only and on the formal use of the
summed form of PT series in a region
where it is not supposed to be valid.
One can think of eq.~(\ref{betamod}) as of a Pade approximation
for the $\beta$ function in a particular scheme.

Some explicitly nonperturbative extrapolations exist also.
The simple pattern is provided by the minimal
analytic continuation
$$
\alpha^{eff}(Q^2) = {1\over ln{Q^2\over \Lambda^2}} - {1\over
{Q^2\over \Lambda^2}-1}
$$
that is regular
everywhere outside a cut in the complex $Q^2$ plane but looks
nonperturbatively in terms of an asymptotic charge
$
\alpha(Q^2) = \left(ln{Q^2\over \Lambda^2}\right)^{-1}
$
$$
\alpha^{eff}(Q^2)=
\alpha(Q^2)-{1\over e^{1\over \alpha(Q^2) }-1}=
\alpha(Q^2)- e^{-{1\over\alpha(Q^2)} }+\ldots
$$
and makes an extraction of ``purely'' nonperturbative terms
untransparent.
The corresponding $\beta$ function
$$
\beta(\xi)= -\xi^2+ (e^{1\over \xi}-1)^{-2}
$$
also has an explicit nonperturbative term unlike eq.~(\ref{betamod}).

Thus, there are some arguments in favor of smooth behavior of the
coupling constant in the IR region keeping positivity.

5. Nonperturbative terms and parameterization.

Going back to our main problem of finding a difference between
$$
\int_0^{Q^2}\alpha_s(t) \omega(t) dt
\quad
{\rm and}
\quad
\alpha_s(Q^2)\int_0^{Q^2} \omega(t) dt
$$
with a newly defined positive $\alpha_s(t)$
one sees that new contributions should be added to the PV
prescription
--
nonperturbative terms have appeared.
It would be convenient to have a clear distinguish between PT and power
corrections that are hidden in the exact formula (\ref{int}).
The simple parameterization could be also useful for
comparison
between
different channels.
We only need our function
defined under the integration sign so we can account
for the nonperturbative
terms by localized distributions living at $t=\Lambda^2$.
We write \cite{bub}
\begin{equation}
\alpha^{eff}(t)=\alpha(t)_{|_{PV}}
+ A\Lambda^2\delta(t-\Lambda^2)+ B\Lambda^4\delta'(t-\Lambda^2)+\ldots
\label{par}
\end{equation}
Parameters $A,B,\dots $ can be found for any
particular model of extrapolation
of running coupling constant into the infrared region.
An important observation is that for a wide class of
extrapolations
an additional constraint  $A=0$ can be satisfied.
It is not necessary but looks plausible because in simple
cases such a condition corresponds to the absence of gauge
invariant operators with the mass dimension two.
In other words we can extend our running coupling without
the large distortion of PT.
Several models belonging to the above class are:

1. No interaction at small momenta
$$
\alpha^{eff}(z)=\alpha(z)\Theta(z-a\Lambda^2).
$$
System of equations for determining the cutoff $a$
and parameters $A$ and $B$ is
$$
li(a)+A=0,\quad li(a^2)-B=0
$$
where $li(a)$ is a special function
$$
li(a)=\int^a_0{dt\over ln (t)}
$$
with the PV
prescription for the pole at real positive
$a>1$.  A solution (with a constraint  $A=0$) is $a=1.45$,
$B=li(2.1)=1.19$.

2. Freezing ({\it e.g.} \cite{schdys})
$$
\alpha^{eff}(s)=\alpha(a\Lambda^2)\Theta(a\Lambda^2-s)
+\alpha(s)\Theta(s-a\Lambda^2).
$$
The system of equations is
$$
{a\over ln (a)}=li(a),\quad{a^2\over 2ln (a)}=li(a^2)-B.
$$
with the solution
$a=3.85,~B=2.6$.

3. Minimal subtraction
\begin{equation}
\alpha^{eff}(s)=({1\over ln(s/\Lambda^2)}
-{1\over s/\Lambda^2-1})\Theta(a\Lambda^2-s)+\alpha(s)\Theta(s-a\Lambda^2).
  \label{minsubmod}
\end{equation}
Again
$$
li(a)-ln(a-1)=li(a),\quad li(a^2)-a-ln(a-1)=li(a^2)-B
$$
and
$a=2,~B=2$.

4. Models generated by the $\beta$ function of eq.~(\ref{betamod})

The solution depends on the parameter $\kappa$ entering the expression for
$\beta$ function (\ref{betamod}).
For $\kappa\sim 2$ the solution is close to one
of models (2) and (3) but for
different $\kappa$ there may be no solution at all
or one with a nonzero $A$.

Thus, for reasonable extrapolations the solution subjected
to the condition $A=0$ does exist and is stable enough,
{\it i.e.} $B$
does not change much.
As for the model (3) one could allow
the formula (\ref{minsubmod})
be valid for any $s$ without any cutoff $a$. Then such an
extrapolation, though quite legal, would be nonminimal in our
sense and distort the PT strongly. It would introduce
$1/Q^2$ terms \cite{alt}
in cases where they seem to be forbidden by
operator product expansion.
But in fact they were just correspond to a
definition of the perturbative
series.

Now one can normalize our parameterization in a
particular case to express the nonperturbative parameter
$B$ through
some known quantities and to predict new ones.
The place to turn for normalization is $e^+e^-$ annihilation
where the operator expansion is known. We find \cite{bub}
\begin{equation}
\langle {\alpha_s\over
\pi}G^2\rangle ={12\over\pi\beta_0}B\Lambda^4.
\label{norm}
\end{equation}
For
$\langle\alpha_s G^2\rangle = (0.440~ MeV)^4$ the
relation between the gluon condensate and $\Lambda$
for this particular model is
$0.330~ MeV
\sim 1.52 B^{1/4}\Lambda$.
For models (2) and (3) we find a constraint
$0.330~ MeV=(1.87\pm 0.18)\Lambda_{\overline{{\rm MS}}}$.
Taken literary it gives
$\Lambda_{\overline{{\rm MS}}}=180\pm 20~MeV$
that is in a reasonable agreement with the present data.
In general, our result means that the numerical values of the gluon
condensate and of
the parameter $\Lambda_{\overline{{\rm MS}}}$ are compatible with each
other for smooth continuation into the infrared region
(like models (2,3)). For the model (1) numerics will be slightly
different and in case of the model (4) it depends on $\kappa$
ranging from a bad solution through the standard values to some bad
ones again.

The generalization to two loop approximation for the asymptotic charge
is straightforward. The Landau pole still exists (though its location
is changed a bit) and the above machinery can be applied.

6. Universality.

Here we assume the hypothesis of universality
\cite{akh,uni,kor}
to predict power
corrections to a
number of observables. We discuss the pole mass and the
event shapes. In the framework of resummation of the
perturbative corrections by a particular method of summation,
the uncertainties, or the limit of accuracy,
for physical observables are normally
discussed.
Our approach, connecting different channels, allows one to
express one observable through another using the
parameterization through the gluon condensate as an intermediate
step
and therefore gives the absolute magnitude of corrections.

First example is the pole mass of a heavy quark.
Being well defined in the PT framework it allows the
representation of the form (\ref{int}) by inserting the bubble
chain into the one loop mass operator. The difference between the
pole mass and the running mass entering the renormalized QCD
Lagrangian is given by \cite{big}
\begin{equation}
m_P - m_Q = {8\pi\over 3}\int_{|k|<\mu}{d^3k\over (2\pi)^3}
{\alpha_s(k)\over {\vec k}^2}
={2\over 3\pi}\int_0^{\mu^2}{\alpha_s(t)dt\over \sqrt{t}}
  \label{pmas}
\end{equation}
Using our parameterization (\ref{par})
we get the result
$$
m_P - m_Q  ={B\over 3\pi\beta_0}\Lambda={4B\over 3b}
$$
at $\beta_0=b/4\pi$
which should be compared with the uncertainty found in \cite{big}
$$
m_P - m_Q  ={8\over 3 b}\Lambda.
$$
The value of the parameter $B$ is determined by the gluon
condensate and is about two from (\ref{norm}).

Second example is the mean value of the
thrust parameter $T$ that has been computed
up to the next-to-leading order in $\alpha_s$
\cite{kun}
$$
\langle T \rangle = 1-0.355\alpha_s(Q^2)(1+9.56{\alpha_s(Q^2)\over \pi}).
$$
Corresponding uncertainties due to bubble chain summation
and estimates of generated power corrections have been
considered
in \cite{man,web,akh}.
We use the representation of the form (\ref{int}) given
explicitly in \cite{web}
\begin{equation}
\delta\langle T \rangle =
-{16\over 3\pi Q}\int_0^Q dk_\perp \alpha_s(k_\perp)
=-{8\over 3\pi Q}\int_0^{Q^2}{\alpha_s(t)dt\over \sqrt{t}}
  \label{wethr}
\end{equation}
Our parameterization leads to the result
$$
\delta\langle T \rangle
=-{16B\over 3 b}{\Lambda\over Q}.
$$
Note, that in this particular example formula (\ref{wethr})
coincides with the
previous case (\ref{pmas}) up to numerical factors.

The dependence on the scheme and specification of $\Lambda$
come from the analyses of the perturbative part of the
expansion.
If the particular
scheme is chosen for PT expansion then the corresponding $\Lambda$
appears in (\ref{wethr}). We are going to elaborate on this point elsewhere.

7. Conclusion.

We propose to parameterize the power corrections in a
well-defined way. The universality hypothesis (physical
justification of which we don't discuss here) allows one to
connect these corrections in different channels.
The PT structure is reflected in the choice of the parameter
$\Lambda$:
as soon as the PT scheme is fixed the corresponding parameter
$\Lambda$
appears in power corrections in our approximation.

AAP is indebted to J.H.Kuhn for kind hospitality
extended to him at University of Karlsruhe where
a part of this paper was written and interest in the work.
AAP thanks K.G.Chetyrkin, M.Jezabek and T.Mannel for discussions,
S.Narison and M.Beneke for correspondence.
Financial support of Soros Foundation
and of RFFR under grant No. 93-02-14428
is gratefully acknowledged.


\end{document}